# Hidden surface photovoltages revealed by pump probe KPFM

*Valentin Aubriet[1], Kristell Courouble[2], Olivier Bardagot[3,4], Renaud Demadrille[3], Lukas Borowik[1], Benjamin Grévin[3]*

[1]Univ. Grenoble Alpes, CEA, LETI, F-38000 Grenoble, France

[2]STMicroelectronics, 850 Rue Jean Monnet, 38926 Crolles Cedex, France

[3]Univ. Grenoble Alpes, CNRS, CEA, INAC-SyMMES, 38000 Grenoble, France

[4]current address: Department of Chemistry and Biochemistry, University of Bern, Freiestrasse 3, Bern, CH-3012, Switzerland

*Corresponding authors: benjamin.grevin@cea.fr , Lukasz.borowik@cea.fr

# Abstract

In this work, we use pump-probe Kelvin Probe Force Microscopy (pp-KPFM) in combination with non-contact atomic force microscopy (nc-AFM) under ultrahigh vacuum, to investigate the nature of the light-induced surface potential dynamics in alumina-passivated crystalline silicon, and in an organic bulk heterojunction thin film based on the PTB7-PC71BM tandem. In both cases, we demonstrate that it is possible to identify and separate the contributions of two different kinds of photo-induced charge distributions that give rise to potential shifts with opposite polarities, each characterized by different dynamics. The data acquired on the passivated crystalline silicon are shown to be fully consistent with the band-bending at the silicon-oxide interface, and with electron trapping processes in acceptors states and in the passivation layer. The full sequence of events that follow the electron-hole generation can be observed on the pp-KPFM curves, *i.e.* the carriers spatial separation and hole accumulation in the space charge area, the electron trapping, the electron-hole recombination, and finally the electron trap-release. Two dimensional dynamical maps of the organic blend photo-response are obtained by recording the pump-probe KPFM curves in data cube mode, and by implementing a specific batch processing protocol. Sample areas displaying an extra positive SPV component characterized by decay time-constants of a few tens of microseconds are thus revealed, and are tentatively attributed to specific interfaces formed between a polymer-enriched skin layer and recessed acceptor aggregates. Decay time constant images of the negative SPV component confirm that the acceptor clusters act as electron-trapping centres. Whatever the photovoltaic technology, our results exemplify how some of the SPV components may remain completely hidden to conventional SPV imaging by KPFM, with possible consequences in terms of photo-response misinterpretation. This work furthermore highlight the need of implementing time-resolved techniques that can provide a quantitative measurement of the time-resolved potential.

# 1. Introduction

For almost two decades, surface photovoltage (SPV) imaging by Kelvin Probe Force Microscopy (KPFM) has been used for local investigations of charge photo-generation, transport and recombination mechanisms in an ever-growing class of photoactive materials.

KPFM is a well-known electrostatic mode of the Atomic Force Microscope (AFM), which yields a measurement of the tip-sample contact potential difference (CPD). The CPD in its most general form[1] is directly related to the difference between the sample and tip vacuum levels (the vacuum level corresponding here to the energy of an electron at rest just outside the solid). This provides a local measurement (relative to the tip) of the sample work function for metals, and gives access to any kind of charge distribution that shifts the energy bands in a semiconductor.

For its part, the SPV can be defined as an illumination-induced changed in the surface electrostatic potential,[2] arising from charge transfers and redistributions. A textbook example is the one of an SPV generation at the surface of a doped-silicon sample, for which band bending at the surface exists in the dark state due to surface states filling. The resulting built in electric field separates the photo-generated electrons/holes across the space charge region. In the ideal case where there are no other in-dark charge distributions (*i.e.* band bending) deeper in the bulk (and no non-uniform generation/recombination mechanisms[2]), the illumination-induced CPD shift probed by KPFM originates only from surface (or more specifically near-surface) contributions. Then, employing the expression surface photovoltage carries no ambiguity.

A well-known fact, however, is that the photovoltage probed at a sample surface can also take its source at buried interfaces. The silicon p-n junction is the archetypal example of such a situation. As stressed by Kronik and Shapira,[2] the surface potential is linked to any illuminated region within the sample where a photovoltage can be generated due to band bending or defect states. Consequently, in some cases, it may become challenging to know ascribe precisely the origin of the observed SPV. In silicon samples, the pioneering KPFM work of Loppacher et al. has nicely illustrated how the surface-related SPV can be dominated by a recessed one.[3] The difficulty raised by their work lies in the fact that in practice, it is impossible to separate the different contributions to the total SPV probed by KPFM.

Since that time, KPFM has been intensively applied to characterize the opto-electronic properties of various families of photoactive materials, displaying each specific photo-generation, transport and recombination mechanisms. There is now a huge literature dealing with SPV measurements by KPFM on silicon, on chalcogenides, on organic donor-acceptor bulk heterojunctions or on hybrid perovskites, to name a few (for a review concerning KPFM on the last two material families, see for instance ref. [4]). However, whatever the material, the question of the origin (*i.e.* surface vs bulk) of the observed photo-potentials remains too often overlooked, and the CPD shift probed by KPFM under illumination is simply referred to as an SPV since it is measured at the sample surface with a scanning probe technique.

The KPFM community has been more prone to concentrate its efforts on correlating nanostructural and/or chemical defects (such as grain boundaries in chalcogenides or perovskites[5,6]) with SPV contrasts in materials that compose the active layers of third generation solar cells. These defects can indeed significantly limit the energy-conversion performances of operating devices, by acting as recombination and/or trapping centres. Moreover, since dynamical processes are at play, the need to develop time-resolved SPV measurements with nanoscale resolution became obvious very early on.

The first attempts to use time-resolved electrostatic modes of the atomic force microscope (AFM) to map the charge dynamics in bulk heterojunctions were made already fifteen years ago.[7] Since, several powerful extensions of the Electrostatic (EFM) and Kelvin probe force microscope (KPFM) have been successfully implemented to monitor dynamical charge and ionic processes at the nanometre scale in organic, inorganic and hybrid photovoltaic materials and devices. [8-21]

Pump-probe Kelvin Probe Force Microscopy (pp-KPFM) is an interesting approach for time-resolved surface potential measurements, since it can be implemented without too much complication on an AFM/KPFM setup. Its basic operation principle, introduced by Muraswki *et al.* a few years ago,[10] consists in restricting the application of the modulated bias voltage ($V_{ac}$) used to detect the electrostatic forces in KPFM to a finite time window. The compensation bias ($V_{dc}$) generated by the KPFM loop matches consequently the contact potential difference (CPD) averaged during this user-defined time-window. Time-resolved measurements are then simply performed by recording curves of the CPD as a function of the delay between a pump signal (which can be an electrical or an optical pulse) and the probe (the "time-windowed" ac modulation bias). For the sake of completeness, one shall note that another pump-probe KPFM approach based on a different operating scheme has been proposed by Schumacher *et al.*.[22]

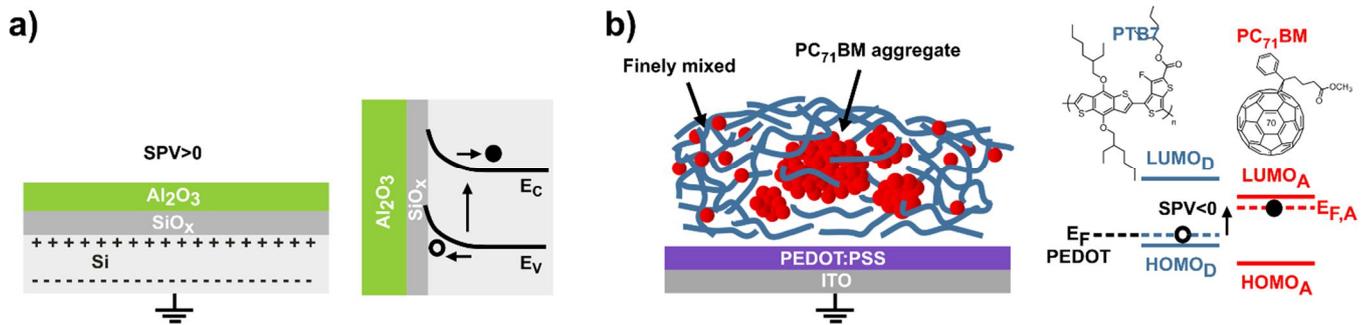

**Figure 1** Samples investigated in this work. **a)** $Al_2O_3$-passivated c-Si. In these samples, one expects a positive surface photovoltage (SPV), due to an upward band bending at the c-Si/oxide interface. Holes accumulate near the film surface under illumination, giving rise to a positive shit of the surface potential. **b)** Left: PTB7-$PC_{71}BM$ blend deposited on an ITO (indium thin oxide) substrate coated by PEDOT:PSS (conducting polymer). Right: donor (PTB7) and acceptor ($PC_{71}BM$) chemical structures, along with the type-II molecular levels alignment. The acceptor and donor species become negatively and positively charged under illumination, respectively. The SPV can in a first approach be related to the quasi Fermi level splitting under illumination. When the D and A species are finely intermixed, one expects a negative surface photovoltage, thanks to the donor Fermi level alignment (pinning) with the substrate (ITO coated by PEDOT:PSS). The case of the samples processed without additive is specific. They feature mesoscopic $PC_{71}BM$ aggregates, partially covered by a PTB7-enriched skin layer.

So far, pp-KPFM investigations of photovoltaic thin films remain extremely confidential. Recently, we have shown that this technique can be used to map 2D images of the SPV dynamics, by implementing a data-cube acquisition mode of the spectroscopic curves[23] (*i.e.* curves of the surface potential as a function of the pump-probe delay). In this work, we further highlight the usefulness of pp-KPFM, by demonstrating its ability to identify different components of the surface photovoltage (characterized each by their own dynamics) in photovoltaic thin films. These investigations were carried out on two kinds of samples (**Figure 1**) selected as model systems for the inorganic and organic photovoltaic communities: p-doped crystalline silicon (c-Si) wafers passivated by aluminium oxide ($Al_2O_3$), and photovoltaic donor-acceptor organic blends based on the PTB7:$PC_{71}BM$ tandem (PTB7: poly[4,8-bis[(2-ethylhexyl)oxy]benzo[1,2-b:4,5-b′]dithiophene-2,6-diyl][3-fluoro-2-[(2-ethylhexyl)carbonyl]thieno[3,4-b]thiophenediyl] , $PC_{71}BM$: [6,6]-Phenyl C71 butyric acid methyl ester). In both cases, it will be shown that the steady state CPD shift (*i.e.* the SPV that would be recorded under continuous wave illumination) originates from two populations of photogenerated charge carriers giving rise to SPV with opposite polarities and different decay dynamics. Our results show that a great caution should be used when analysing the SPV data obtained by conventional KPFM, since "hidden" SPV components can only be revealed by time-resolved measurements. They also point out the need to use time-resolved modes of the AFM that yield a truly quantitative measurement of the surface potential.

## 2. Methods

### 2.1 pp-KPFM setup

Pump-probe KPFM (**Figure 2**) investigations were carried out with a non-contact AFM beam-deflection setup operated under ultra-high vacuum at room temperature, and driven by a Matrix control unit (VT-AFM, Scienta Omicron).

In this work, the time-resolved KPFM loop is based on a heterodyne-KPFM[24-26] scheme (**Figure 2a**). This mode is similar to side-band KFPM, for which the electrostatic interaction is detected at the sidebands $\omega_0 \pm \omega_{mod}$ of the mechanical oscillation. ($\omega_0$ and $\omega_{mod}$ stand for the frequencies of the mechanical oscillation at the first resonance mode and of the electrical modulation, respectively).

In heterodyne-KPFM, the sideband frequency is shifted at the second eigenmode (frequency $\omega_1$) by performing the electrical excitation at $\omega_{mod}=\omega_1-\omega_0$. Doing so, one combines the advantages of amplitude modulation KPFM (AM-KPFM) and frequency modulation KPFM (FM-KPFM). Likewise to conventional AM-KPFM, the sensitivity is boosted by performing the amplitude detection at the second resonance, but heterodyne KPFM is not prone to stray capacitance effects.[24] Indeed, the signal used for electrostatic detection in heterodyne-KPFM is proportional to the force gradient[26] (through the second z-derivative of the tip-sample capacitance). Heterodyne KPFM was implemented by employing a MFLI digital lock in amplifier from Zurich Instruments. In situ annealed Pt/Ir-coated silicon cantilevers were used for all experiments (EFM, nanosensors, mechanical resonance frequency around 75kHz, second eigenmode around 475kHz).

The pump and probe pulse signals are generated by a dual channel arbitrary waveform generator (AWG, Keysight Instruments, 33622A), following pre-programmed pulse-probe delay sequences[23]. The pump- probe sequences are triggered by the scanning probe unit, which allows recording pp-KPFM curves either at selected locations of the sample or onto a 2D grid. The pump pulses drive a digital modulation laser source (PhoxPlus laser module from Omicron laserage, λ=515nm), and the "time-windowed" KPFM signal is obtained by multiplying the probe pulses (**Figure 2b,c**) by the heterodyne KPFM loop potentials (the sum of the ac modulation and the dc compensation potentials). This is achieved by mean of a dedicated circuit based an analog multiplier (AD835, Analog devices).

This analog stage also includes a summation module (built around an AD8130 differential amplifier from Analog devices), that is used to sum the time-windowed potentials (generated by the

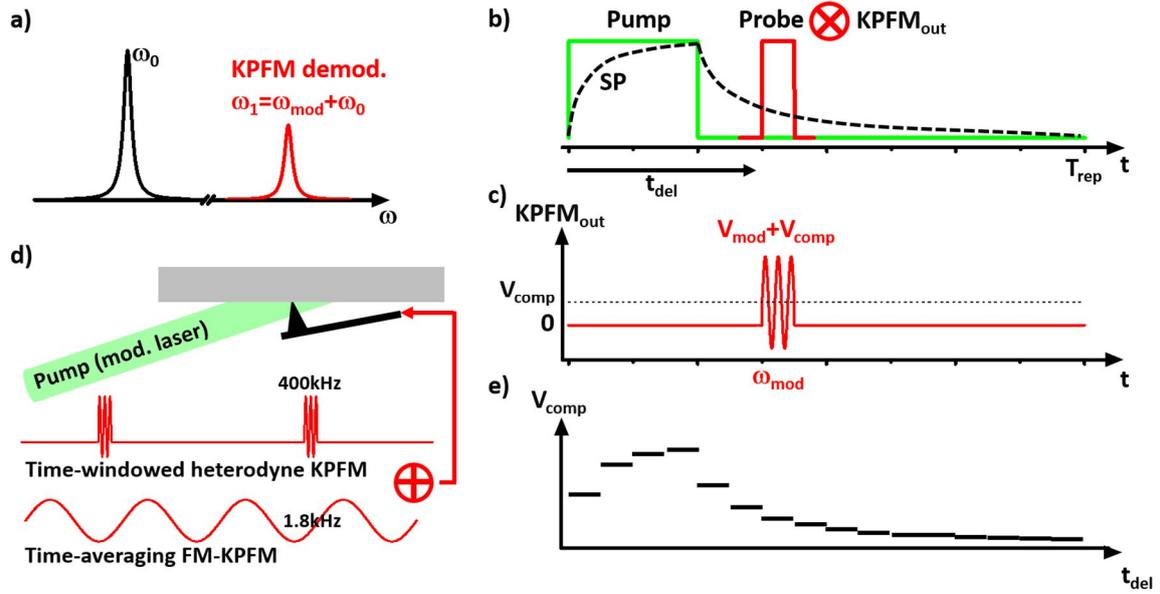

**Figure 2** pp-KPFM implementation. **a)** The time-resolved KPFM loop is based on an heterodyne detection. The KPFM controller tracks variations is the response amplitude of the second cantilever eigenmode (frequency $\omega_1$), which is excited by applying a modulation bias voltage ($V_{mod}$) at a frequency $\omega_{mod}$ equal to $\omega_1$-$\omega_0$. A modulated component of the electrostatic force at $\omega_1$ is generated by the interaction of the tip oscillation (first resonance mode at $\omega_0$) with the oscillating electric field ($\omega_1$-$\omega_0$). **b)** A dual output waveform generator is used to generate synchronized pump-probe periodic sequences (repetition period $T_{rep}$). The pump pulses drive the input of a digital modulation laser source that illuminates the sample. The probe pulses are mixed with the bias output of the KPFM controller. **c)** The modulated bias voltage ($V_{mod}$, $\omega_{mod}$) is only applied during the time-window defined by the probe pulses. Consequently, the KPFM compensation bias ($V_{comp}$) that minimizes the demodulated electrostatic force (at =$\omega_1$) yields a measurement of the time-averaged surface potential over the probe time-window. Note that $V_{comp}$ is also only applied during the probe time-window. **d)** A second KPFM loop operated in frequency modulation mode (FM-KPFM) is added to avoid artefacts in the z-regulation, due to non-compensated electrostatic forces outside the probe time-window. The outputs of the "time-windowed" heterodyne KPFM loop and FM-KPFM loop (modulation bias and compensation bias) are summed and sent to the tip. For small duty ratios, the time-resolved surface potential is equal to the sum of the two compensation potentials (see main text). **e)** Time-resolved measurements are performed by recording curves of the KPFM compensation potential as a function of the pump-probe delay $t_{del}$.

time-resolved heterodyne KPFM loop) with the output of a second KPFM loop (**Figure 2d**), operated in standard FM-KPFM (on the basis of a 7280 dual harmonic lock-in amplifier from Signal Recovery). This conventional loop compensates the time-averaged electrostatic potential.

As explained by Murawski *et al.*,[10] this prevents the occurrence of topographic artefacts that may arise due to time-variable electrostatic forces (since the time-resolved loop does not compensate the time-average electrostatic potential). Note that in our setup, the demodulations are performed via different channels (frequency shift of the first mode and amplitude of the second eigenmode for FM-KPFM and heterodyne-KPFM, respectively), and rely on two modulated voltage bias located far away from each other in the frequency domain (typ. 1.8kHz for the FM-KPFM loop and 400kHz for the heterodyne-KPFM loop). This ensures that the two KPFM detection paths are fully independent. Nevertheless, the two compensation voltages are simultaneously applied during the probe time-window. Consequently, in the dual loop scheme, a post processing calculation is needed to extract the time-resolved and the time-averaged surface potentials. It can be shown that for small duty ratios (i.e. the probe time-window divided by the pump-probe sequence repetition period), the time-resolved potential is given by the sum of both loops compensation potentials.[10]

The pump-probe pulse sequences were programmed such a way that each pump-probe delay is repeated several times before moving to the next value. This procedure insures that enough time is given to the time-resolved KPFM loop to properly track the SP change that occur when moving the pump–probe delay.[23] Only the last data points were kept for each delay, the number of repetitions are indicated in the Figure captions (additional details can be found in the supplementary information, see Figure S1)

Last, sample illumination was performed with the modulated laser source through an optical viewport of the AFM setup. Organic heterojunctions (deposited on transparent substrates) were illuminated in backside configuration.[23] For each measurement, the optical power (Popt corresponding to the maximum pulse intensity during the modulated illumination) is indicated in the corresponding figure caption. Popt is defined per unit of surface by taking into account the laser beam diameter.

## 2.2. pp-KPFM curves adjustment

Once the pp-KPFM data have been acquired, a mathematical adjustment is needed to extract the time constants that characterize the surface photovoltage dynamics. **Figure 3** presents a scheme of the surface potential time-evolution under pulsed illumination for one given pump-probe delay (Δt). The time-intervals corresponding to the pump and probe signals are highlighted by half-transparent green and red rectangles. To avoid any confusion, it is important to note that the time (t) and pump-probe delay (Δt) scales coincide only because the delays are defined with respect to the time origin t=0. The experimental data discussed in this work will be presented with a single x-scale (pp-KPFM potential as a function of Δt). In the following, it is important to keep in mind that each pp-KPFM data point represents a measurement of the surface potential integrated during the probe time-window.

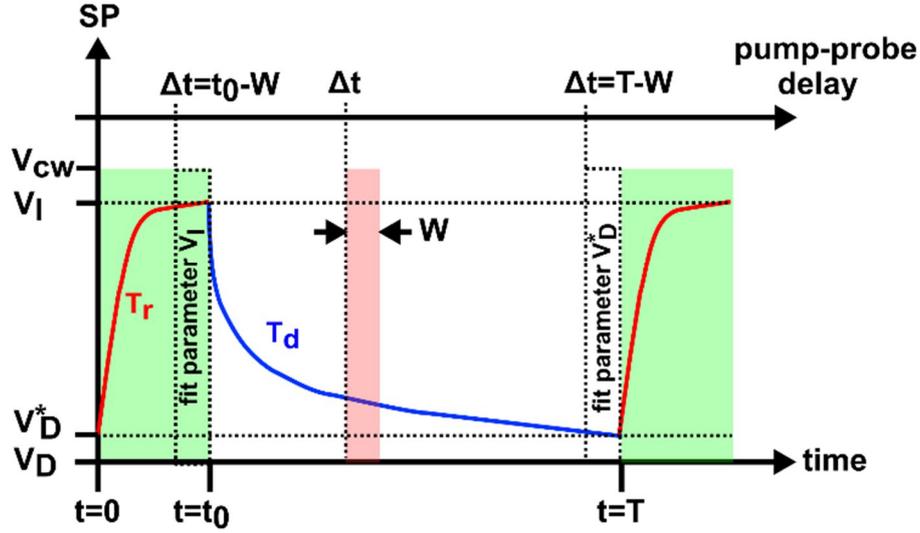

**Figure 3** Surface potential (SP) time-evolution under pulsed illumination, along with the pump and probe pulses. Half-transparent green and red rectangles show the respective time-positions of pump and probe signals, for a given delay (Δt). The illumination occurs for 0≤t≤$t_0$. w: probe time-window. T=pump/probe sequence period. $V_D$: surface potential in dark. $V_I$: maximum value reached by the surface potential under illumination (case of a positive photovoltage). If the pulse duration ($t_0$) exceeds largely the photovoltage rise time constant ($\tau_r$), $V_I$ equals the SP value that would be measured under continuous wave illumination ($V_{cw}$). $\tau_d$: photovoltage decay time-constant. If T-$t_0$ exceeds largely $\tau_d$, the surface potential at t=0 ($V_D^*$) equals the in dark surface potential ($V_D$). For the curve adjustment, $V_I$ and $V^*_D$ parameters are fixed to the values of the pp-KPFM potential measured for Δt=$t_0$-w and Δt=T-w (see main text).

First, we derive the formula used to fit the "decay part" (Δt≥$t_0$) of pp-KPFM spectroscopic curves. The surface photo-voltage (SPV) decay dynamics between t=$t_0$ and t=T (see Figure 2) can be modelled by using a stretched exponential function (with time constant $\tau_d$ and stretch exponent β):

$$SP(t) = (V_I - V_D) \times e^{-\left[\frac{t-t_0}{\tau_d}\right]^\beta} + V_D \quad (1)$$

$V_I$ stands for the maximum (or minimum depending on the SPV polarity) value reached by the surface potential at the end of the light pulse (t=$t_0$), and $V_D$ for the in-dark surface potential. $\tau_d$ is a time-constant that characterizes the SPV decay dynamic. If relevant, dispersive kinetics can be accounted by the stretch exponent (otherwise β is fixed to unity).

As highlighted before, the pp-KPFM potential corresponds to the time-averaged value of SP(t) over the probe time window (Δt≤t≤Δt+w, see **Figure 3**):

$$\frac{1}{w}\int_{\Delta t}^{\Delta t+w} SP(t)dt \quad (2)$$

It follows[23] that the pp-KPFM potential for a given pump-probe delay Δt is:

$$pp-KPFM^{decay}(\Delta t) = (V_I - V_D) \times \frac{\tau_d}{w \times \beta} \times \left(\gamma\left[\frac{1}{\beta}, \left(\frac{\Delta t+w-t_0}{\tau_d}\right)^\beta\right] - \gamma\left[\frac{1}{\beta}, \left(\frac{\Delta t-t_0}{\tau_d}\right)^\beta\right]\right) + V_D \quad (3)$$

In this last formula, γ stands for the non-normalized lower half Euler Gamma function.

The formula used to adjust the "rise part" ($0 \leq \Delta t \leq t_0$) of the spectroscopic curves can be derived in a similar fashion. Again, a stretch-exponential based function is used to model the photocharging dynamics (characterized by a photopotential rise time constant $\tau_r$):

$$SP(t) = (V_I - V_D^*) \times \left(1 - e^{-\left[\frac{t}{\tau_r}\right]^\beta}\right) + V_D^* \quad (4)$$

It is important to note that the surface potential at t=0 (beginning of the pump-probe sequence) does not necessarily equals the one that would be measured in dark conditions ($V_D$). Indeed, such an initial state exists only if the time interval between the light pulses exceeds largely the SPV time-decay (see **Figure 3**). For that reason, the "initial" surface potential SP(t=0) is referred to as $V^*_D$, and the difference $V_I - V^*_D$ yields a "pseudo SPV" amplitude (by reference to the full SPV amplitude, $V_I - V_D$). Again, the pp-KPFM potential is obtained by integrating the surface potential over the probe time-window:

$$pp-KPFM^{rise}(\Delta t) = (V_I - V_D^*) \times \left[1 - \frac{\tau_r}{w \times \beta} \times \left(\gamma\left[\frac{1}{\beta}, \left(\frac{\Delta t+w}{\tau_r}\right)^\beta\right] - \gamma\left[\frac{1}{\beta}, \left(\frac{\Delta t}{\tau_r}\right)^\beta\right]\right)\right] + V_D^* \quad (5)$$

So far, we made no assumption with regard to the SPV polarity; both formulas (*i.e.* Equ. 3 and Equ. 5) can be applied whatever the direction of the light-induced potential shift. However, as shown hereafter, the global SPV probed by KPFM sometimes results from the cumulative contributions of two kinds of light-induced charge redistributions in the sample, which give rise to potential shifts with opposite polarities.

From an electrostatic point of view, such a situation can be understood by adding in series two potential drops. Generalized formulas can therefore simply be obtained for the SPV rise and decay regimes by summing two terms characterized by their own amplitudes and dynamical parameters:

$$pp-KPFM^{decay}(\Delta t) = (SPV_+) \times \frac{\tau_d^+}{w \times \beta^+} \times \left(\gamma\left[\frac{1}{\beta^+}, \left(\frac{\Delta t+w-t_0}{\tau_d^+}\right)^{\beta^+}\right] - \gamma\left[\frac{1}{\beta^+}, \left(\frac{\Delta t-t_0}{\tau_d^+}\right)^{\beta^+}\right]\right) + (SPV_-) \times \frac{\tau_d^-}{w \times \beta^-} \times \left(\gamma\left[\frac{1}{\beta^-}, \left(\frac{\Delta t+w-t_0}{\tau_d^-}\right)^{\beta^-}\right] - \gamma\left[\frac{1}{\beta^-}, \left(\frac{\Delta t-t_0}{\tau_d^-}\right)^{\beta^-}\right]\right) + V_D \quad (6)$$

$$pp - KPFM^{rise}(\Delta t) = (SPV_+^*) \times \left[1 - \frac{\tau_r^+}{w \times \beta^+} \times \left(\gamma\left[\frac{1}{\beta^+}, \left(\frac{\Delta t+w}{\tau_r^+}\right)^{\beta^+}\right] - \gamma\left[\frac{1}{\beta^+}, \left(\frac{\Delta t}{\tau_r^+}\right)^{\beta^+}\right]\right)\right] +$$

$$(SPV_-^*) \times \left[1 - \frac{\tau_r^-}{w \times \beta^-} \times \left(\gamma\left[\frac{1}{\beta^-}, \left(\frac{\Delta t+w}{\tau_r^-}\right)^{\beta^-}\right] - \gamma\left[\frac{1}{\beta^-}, \left(\frac{\Delta t}{\tau_r^-}\right)^{\beta^-}\right]\right)\right] + V_D^* \quad (7)$$

In both formulas, the first and second group of terms account for the contributions of photovoltages with positive and negative polarities. The sum of their amplitudes must be equal to the total photovoltage. This condition is not explicitly expressed in the above equations, but it is mandatory to impose linear constraints on the variable parameters when performing the data adjustment:

$$SPV_+ + SPV_- = SPV_{total} = V_I - V_D \quad (8)$$

$$SPV_+^* + SPV_-^* = V_I - V_D^* \quad (9)$$

Here again, the case of the photocharging regime (Equ. 9) differs slightly from the one of the decay regime, due to the possible deviation of the surface potential at t=0 with respect to the in-dark value (*i.e.* V*$_D \neq$V$_D$). For that reason, the SPV component amplitudes have been labelled with stars indices in Equ. (7) and (9), since they are defined relatively to V$_D$* instead of V$_D$.

Last, in addition to the linear constraints on the SPV components amplitudes, the following rules have been applied for the data adjustment. When fitting the data in the decay regime, V$_D$ is a variable and V$_I$ is a fixed parameter which value is fixed to the pp-KPFM potential measured at the end of the pump light pulse (Δt=t$_0$-w, highlighted by a dotted rectangle in **Figure 3**). In turn, V$_I$ becomes a variable when performing a data adjustment in the photocharging regime, and V*$_D$ is fixed to the value reached at the end of the sequence (Δt=T-w, **Figure 3**). Last, in this work, all stretch exponents have been fixed to 1 to limit the number of variable parameters.

## 2-3. Al$_2$O$_3$-passivated c-Si

To further improve the efficiency of silicon-based solar cells, it is vital to minimize (ideally eradicate) the losses that occur by recombination at the wafers surfaces. In that frame Al$_2$O$_3$ appears to be a very effective passivation material for p-type c-Si due a low defect density (~$10^{11}$ eV$^{-1}$cm$^{-2}$) and a high density of fixed negative charges (>$10^{12}$ cm$^{-2}$).[27] Consequently, under illumination one expects a positive value of the SPV caused by the diffusion of holes toward the surface,[28] as shown in **Figure 1a**. On the other hand, several groups report the presence of charge trapping sites in thermal Al$_2$O$_3$ based silicon passivation stack.[29,30] As reported by Kronik *et al*, trapping mechanisms may disturb the surface potential and compete with the diffusion of free carriers.[2] In brief, the SPV signal from Al$_2$O$_3$ passivated c-Si materials can be influenced by two different components (carrier diffusion, trapping mechanisms). In that context, we choose Al$_2$O$_3$ passivated c-Si materials as a model sample for pp-KPFM measurements.

In this work, we deposed 13nm thick Al$_2$O$_3$ over boron-doped (100) oriented 300 mm diameter CZ silicon wafer (N$_D$ = 10$^{15}$ at/cm$^3$). Before Al$_2$O$_3$ deposition, a chemical silicon oxide was obtained using a liquid phase HF standard clean 1 (HFSC1).[31] The Al$_2$O$_3$ layer were deposed at 100°C by thermal atomic layer deposition (ALD) using Al(CH$_3$)$_3$ as precursor and water vapor. Finally, the sample were annealed at 400 °C for 120 minutes under nitrogen.

### 2.4. PTB7:PC$_{71}$BM bulk heterojunctions thin films

Organic bulk heterojunction (BHJ) solar cells are based on the use of complementary electron donor and acceptor π-conjugated materials (semiconducting polymers and/or small molecules), which molecular orbital levels display a type II alignment (see **Figure 1b**) that promotes the exciton dissociation into free-charge carriers (through an intermediate Coulomb-bound charge transfer state). The donor and acceptor species should moreover form nano-phase segregated bi-continuous networks to minimize the exciton losses (by geminate recombination), and convey efficiently the charges to the electrode devices. Introducing in more details the field of organic photovoltaics would be beyond the scope of this work (we refer the reader to review articles[32]). However, it is important to keep in mind that the above picture is far from capturing the complexity of BHJs. For instance, hierarchical organization[33] or three phases[34] models have been proposed to account for the morphology, phase composition and opto-electronic properties of BHJ layers used in operating devices.

PTB7:PC$_{71}$BM donor-acceptor (D-A) blends have provided a platform for basic research in organic photovoltaics (OPV) over several years. Although this D-A tandem does no longer figure among the most efficient ones, we selected it as a model system for our pp-KPFM investigations. It is indeed easy to tune the morphology and phase composition of PTB7:PC$_{71}$BM blends during their solution processing. "Optimized" nanoscale phase segregated samples can be obtained from solutions containing a small amount of 1,8-diiodooctane (DIO).[35-38] Conversely, samples processed without additives feature large-size PC$_{71}$BM aggregates embedded in a finely intermixed donor-acceptor matrix. It has moreover been shown that the acceptor clusters are likely to be surrounded by a polymer-enriched shell, and that the intermixed D-A phase forms a skin layer that covers (at least partially) the film surface.[36] Such samples provide a benchmark to test what is the impact of recessed interfaces to the SPV signals probed by pp-KPFM on organic BHJs.

For this work, a PTB7:PC71BM thin film has therefore been processed from a chlorobenzene (CB) solution without additive, in order to obtain a model system displaying mesoscopic features for pp-

KPFM experiments. A reference sample has also been prepared from a (CB:DIO) mixture (3 volume%). Both films were deposited by solution-casting on ITO substrates coated with PEDOT:PSS (a conducting polymer). PTB7 (purchased from Ossila) was dissolved in CHCl3 and re-precipitated in MeOH before filtration. $PC_{71}BM$ (purchased from Solenne BV, with a 99% purity) was used as received. A thin layer of filtered (0,45µm) PEDOT:PSS (Baytron A14083, Clevios) was spin-coated onto activated ITO surface at 5000rpm/25sec, 4000rpm/60sec/4000rpm.sec-1 (~30 nm) and annealed at 120°C/10min under ambient conditions. The blend deposition was performed into an argon-filled glovebox. PTB7:$PC_{71}BM$ solutions (1:1,5 weight ratio, 25mg/mL total concentration, in anhydrous chlorobenzene) were stirred overnight at 50°C for complete dissolution, and cooled down to RT.

# 3. Results

### 3.1. $Al_2O_3$-passivated c-Si

**Figure 4a** shows the results of the pump-probe KPFM acquisition performed on the Al2O3-passivated c-Si sample. The pump and probe width were set to 2 ms and 250 µs over a 10 ms repetition period, respectively. Forty equally spaced pump-probe delays were used to cover the full repetition period. These parameters were set with the aim to track the full surface potential (SP) dynamics.

In short, these data reveal the existence of four distinct dynamical regimes (highlighted by labels in **Figure 4a**). First, the SP undergoes a positive shift at the beginning of the light pulse. The maximum SP level is almost completely reached at the first pump-probe delay, which indicates that the SPV rise time constant must be on the order or shorter than the probe duration. Strikingly, this fast SP increase (1) is followed by a slower decrease (2) towards a less positive value under illumination. The evolution of the time-resolved signal after the light pulse extinction is even more remarkable. Since the light-induced SPV is positive, one expects an SP decrease towards a more negative (or less positive) in-dark level. A fast decrease occurs indeed (3) for the first three delays that fall outside the pump time-window, however, this drop is followed by a slower increase (4) toward a more positive value. In overall, as illustrated by the inset in **Figure 4a**, this sequence of events can only be explained on the account of the coexistence of positive and negative photo-potentials with different dynamics. Even without a fit, one guess easily that the positive component features smaller dynamical time-constants.

A quantitative analysis of the SPV dynamics has been carried out by mean of mathematical adjustments, performed on the data points corresponding to the photopotential rise (0≤t≤2ms) and decay (2ms≤t≤10ms) regimes. We applied the formulas derived above for the case of a "dual component" photovoltage (Equ. 6 and Equ. 7). The outputs of the fits are reported in **Figure 4b.** A remarkable agreement is achieved between the photopotential amplitudes deduced from the

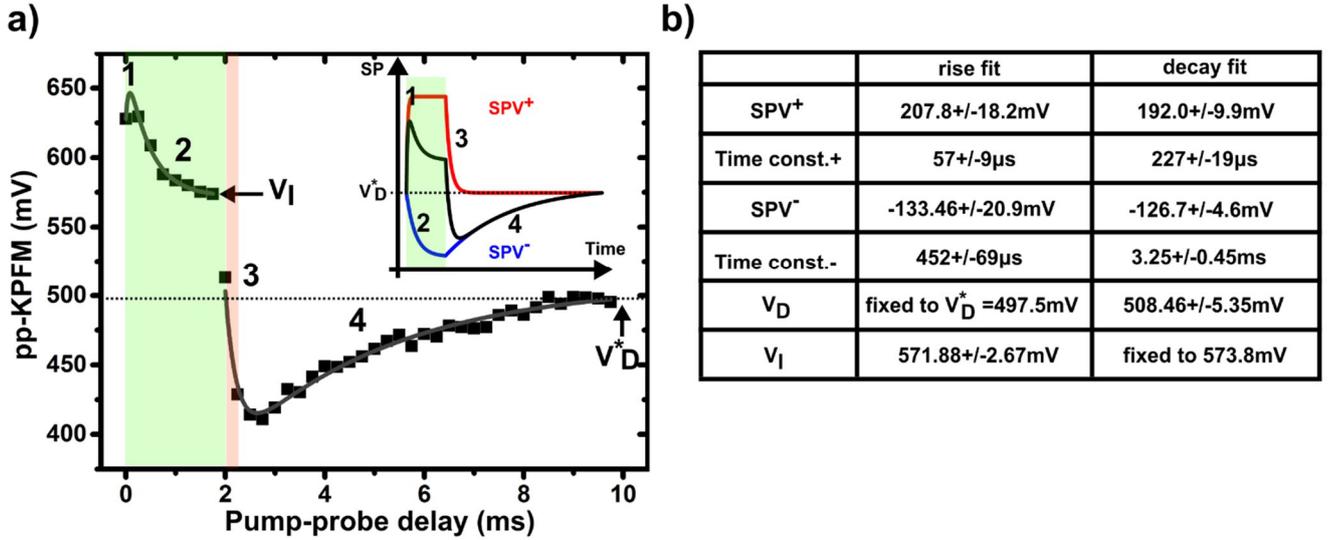

**Figure 4 (a)** Time-resolved pp-KPFM single point spectroscopy on the $Al_2O_3$-passivated c-Si sample. $\lambda$=515nm, Popt=90mW/cm$^2$. Pump duration 2000µs. Probe time-window 250µs. Repetition period 10ms. The measurement has been repeated twice for each pump-probe delay, only the last acquisition has been kept. Half-transparent green and red rectangles show the respective time-positions of pump and probe signals for $\Delta(t)$=2ms. The labels 1 to 4 highlight the positive SPV rise, the negative SPV rise, the positive SPV decay and the negative SPV decay regimes, respectively. Curve adjustments are plotted in dark-grey (continuous lines). For the fits, $V_I$ and $V^*_D$ (indicated by arrows and by a dotted line for $V^*_D$) values have been fixed from the experimental data (see main text). Inset: scheme showing the time-evolution of the positive photovoltage (SPV$^+$ in red), negative photovoltage (SPV$^-$ in blue) and total surface potential SP (in dark). **(b)** Outputs from the data adjustments in the rise and decay regimes.

adjustments of the rise and decay regimes. We stress here that the positive and negative photovoltage amplitudes (SPV+ and SPV-) are true variable parameters in both fits, and that no constraints have been imposed to limit their values (except the rule that imposes an equality between the total photovoltage and the sum of the negative and positive amplitudes). A fairly good agreement is also obtained between the experimental data, and the adjusted values for the in-dark potential and the potential reached at the end of the light pulse $V_I$. It should be here remembered that when one of these last two parameters is fixed during the fit, the other is a variable, and that the roles are reversed when one moves from the adjustment of the rise regime to the one of the decay. We also remind that for the "rise fit", $V^*_D$ shall be used instead of $V_D$ (see the methods). The $V_D$ value extracted from the adjustment of the decay regime is barely higher (*i.e.* more positive by a few mV within the error bar) that $V^*_D$. This confirms that the pump-probe sequence period is long enough to allow a quasi-complete return of the SP to its in-dark state in between the light pulses. To sum it up, the full data set deduced from the two independent adjustments is remarkably self-consistent, which supports the validity of our two-component model. In the following, we will see how the "fast" and "slow" sample photoresponses can be accounted for.

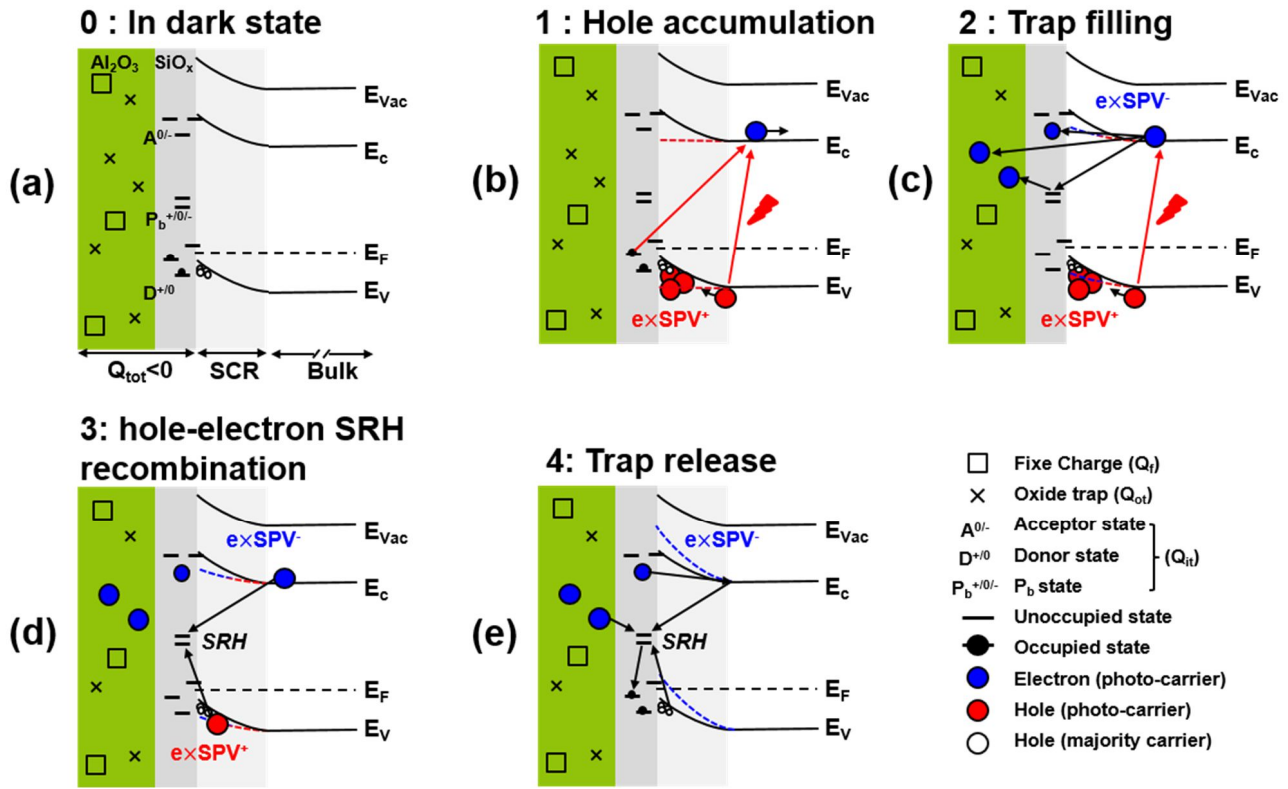

**Figure 5** Simplified band diagram and electrostatic state of the $Al_2O_3$-passivated crystalline silicon sample in dark conditions and under illumination. **(a)** Initial in-dark surface potential configuration. The upward band-bending results from the existence of a permanent total negative charge contained in the passivation stack ($Q_{tot}<0$). The total charge is the sum of sub-components such as fixe charge ($Q_f$), oxide trap ($Q_{ot}$) and interface charge ($Q_{it}$). **(b, c)** Under illumination, the SP increases towards more positive values (step 1, positive SPV rise) due to the accumulation of photo-generated holes (drift under the build in electric field), possibly accompanied by a partial depopulation of donor states. Next, the SP decreases (step 2, negative SPV rise) due to electron-trapping mechanisms which tend to screen the accumulated holes. The red/blue dotted lines illustrate the equilibrium between the positive and negative components. **(d, e)** After the light pulse extinction, photo-generated carriers recombine after a reverse drift (step 3, positive SPV decay), the recombination is assisted by the mid-gap states (SRH). Finally, trapped charges are remitted in silicon (step 4, negative SPV decay) and the SP converges toward the initial in-dark state. Note that the KPFM tip is positioned above the 13 nm thick $Al_2O_3$, and that the photo-carriers are generated in the silicon. For the sake of simplicity, the quasi Fermi level splitting under illumination and the oxide trap energy position within $Al_2O_3$ are not depicted.

A basic understanding of the in-dark sample electrostatic state is an essential prerequisite to the SPV analysis. In the dark state the surface potential results from an equilibrium between all the charges contained in the structure, as shown by **Figure 5a**. In the $SiO_2/Al_2O_3$ stack we distinguish the fixed charges ($Q_f$), the oxide trapped charge ($Q_{ot}$) and the interface charges ($Q_{it}$).[39] The first charge type is related to the oxidation process (ambient oxidation) and is not in electrical communication with the underlying silicon. On the contrary, oxide trapped charges can communicate with silicon through trap/phonon assisted tunneling or charge hopping via defects. Finally, the interface trapped charges are located at the Si-SiO$_2$ interface and can communicate directly with the underlying silicon. We note that the states below (above) the intrinsic Fermi level (not shown here) are referenced to donors (acceptors) states which are positively charged (neutral) when empty and neutral (negatively charged) when occupied.[40] Accordingly, states below the Fermi level are occupied and hence neutral. Among these states, we differentiate the recombination centres and the traps. Precisely, recombination centres are located closer to the mid-gap and can communicate with both bands. The density of states near mid-gap is controlled by the density of dangling bounds also referenced as Pb centre.[41] On the contrary, centres close the bands are identified as traps and can exchange charges to a single band.[39]

In order to respect the charge neutrality principle, under thermal equilibrium condition, the silicon surface is under accumulation due to a permanent negative charge ($Q_{tot}<0$) contained in the SiO$_2$/Al$_2$O$_3$ stack ($Q_{tot}$ corresponds to the sum of $Q_f$, $Q_{ot}$ and $Q_{it}$). In other words, the silicon bands display an upward band bending (**Figure 5a**). Under illumination, the absorption of a photon with an energy higher than the bandgap induces the generation of an electron-hole pair in the silicon. Due to the positive band bending and the associated build in electric field, holes experience a drift toward the surface, which tend to decrease the potential barrier at the Si-SiO$_2$ interface (**Figure 5b**). The observed positive SP shift under illumination is thus fully consistent with the upward band bending induced by the negative permanent charge $Q_{tot}$.

Obviously, under illumination, the charge density at the silicon surface deviates from the equilibrium state. This charge deviation can be described in terms of quasi-Fermi levels (not shown here). Accordingly, due to the deviation of the Fermi level at the silicon surface, some donor states may become unoccupied under illumination which may increase the contribution of the positive charges in the total charge balance. In overall, we thus attribute the rapid increase of the SP to the accumulation of photo-generated holes, possibly accompanied by a partial donor states depopulation (**Figure 5b**). This fast positive SPV rises with a time constant a few tens of μs ($\tau^+_r \sim 57\mu s$). Since the hole accumulation is forced by the internal electric field, this time constant is directly related to the charge transit time (under electric-field-induced drift) that immediately follows the photo-generation.

Next, a decrease of the SP is observed under illumination (step 2 in **Figure 4a**), and finally the SP reaches an equilibrium level (still under illumination). This highlights the presence of slow states that can interact with the positive photo-generated charges. To be more precise, due to the decrease of the surface barrier during the first phase we expect photo-generated electrons to be trapped either in the acceptors states[42] or in the Al$_2$O$_3$ through tunnelling[43] or via charge hopping through mid gap defects.[30] This trap filling process is depicted in **Figure 5c**. The trapped charges increase the negative charge contribution to the total charge balance, resulting in a SP decrease under illumination (*i.e.* a negative SPV). Based on the data adjustment, this charge trapping process occurs within a few hundreds of µs ($\tau^-_r \sim 452$µs).

After the light pulse extinction, the photo-generated electrons can recombine with holes through Shockley–Read–Hall (SRH) recombination. Precisely, we associate the fast decay of the positive SPV component (step 3 in **Figure 4a**) to SRH recombination mechanisms (depicted in **Figure 5d**), which take place at the time-scale of a few hundreds of µs ($\tau^+_d \sim 227$µs). We must highlight that the carrier lifetime describes a property of a carrier within a semiconductor rather than the property of the semiconductor itself.[39] Previous studies on Al$_2$O$_3$ samples, however, report carrier lifetimes in semiconductors ranging from a few hundred µs[44] to a few ms [45] rendering our estimation consistent with the literature.

However, as already mentioned, the SP does not return immediately to its initial in-dark value after SRH recombination. The negative charge carrier trapped in the SiO$_2$-Al$_2$O$_3$ stack are at the origin of this phenomenon. To recover the initial electrostatic landscape, the trapped charges must be remitted in the underlying silicon (**Figure 5e**). Once remitted, they can recombine with a counter charge through SRH mechanism. This de-trapping mechanism (step 4) is a slow thermally- assisted process; the negative SPV component needs a few ms to vanish completely ($\tau^-_d \sim 3$ms).

To summarize this section, we have shown that the total surface photovoltage originates from the cumulative contributions of a photo-induced charge separation across the space charge region (*i.e.* hole accumulation near the silicon surface), and of a negative charge accumulation within defects (traps) in the passivation layer. These processes results in the apparition of positive and negative SPV components, characterized by different rise and decay dynamics. We have provided an unprecedented experimental demonstration, which shows that pp-KPFM can differentiate these processes and their dynamics. In the future, it shall be possible to improve our understanding of the charge transfer mechanisms between the silicon and the passivation stacks, by extending these investigations to a broader variety of passivation materials.

### 3.2 PTB7:PC71BM bulk heterojunctions

The surface morphology of the donor-acceptor blends processed for this work were found to be fully consistent with the ones observed by others for samples processed in the same conditions.[36,38,46] The films processed without solvent additive feature large domains (100-400nm in diameter), as shown in **Figure 6a**. Conversely, the topographic images obtained on the sample processed with DIO additive display a rather uniform contrast, indicating that a finer intermixing of the donor and acceptor species has been obtained (see **Figure S2** in the supplementary information).

Before delving into the time-resolved measurements, it can be helpful to characterize the blend electrostatic landscape and photo-response by "conventional" FM-KPFM. To that end, we performed a data-cube acquisition of the KPFM compensation potential VKPFM recorded in open z-loop as a function of time t ($V_{KPFM}(t)$ curves), the acquisition being synchronized with illumination sequences. A single light pulse is applied during each sequence, as shown in **Figure 6b**. This procedure, which we will call data-cube differential SPV imaging, allows mapping 2D images of the sample's surface potential in the dark state, under continuous wave illumination (provided that a steady state is reached under the light pulse duration), and finally to recalculate SPV images from their difference.[23,47]

In average, the "in dark" surface potential (**Figure 6c**) is more negative over these domains, which is consistent with the existence of large-scale phase separated $PC_{71}BM$ aggregates. At this stage, we just mean that a higher $PC_{71}BM$ concentration shall give rise to a specific KPFM contrast. The "in-dark" electrostatic contrasts reflect the existence of permanent charge distributions,[4] which can either originate from unevenly distributed electrostatic dipoles at the donor-acceptor interfaces,[48] or from permanently trapped charge carriers. Obviously, if the blend composition ratio or the morphology of the donor and acceptor sub-networks vary, the permanent charges (interface dipoles or trapped carriers) shall vary as well. In our case, a plausible hypothesis could be that negative charge carriers remain permanently trapped in the $PC^{71}BM$ clusters, giving rise to a more negative SP.

Taken solely, that model is however insufficient to account for the overall SP contrast. The topography-SP correlation is indeed far from being perfect, clear SP fluctuations are observed within the domains. This already confirms that the $PC_{71}BM$ clusters are in part covered by a PTB7-enriched skin layer. The overall SP contrast can be tentatively understood by taking into account the cumulative contributions of negative trapped charge carriers in the recessed aggregates, and dipoles (which can fluctuate in magnitude and orientation) at the aggregates-skin layer interfaces.

As expected, the peculiar blend morphology impacts the photoresponse, as probed by the SPV imaging (**Figure 6d**). The SPV is more negative in average over some of the acceptor aggregates, which is consistent with the photo-carrier separation mechanism across the D-A interfaces (photo-generated electron and holes are respectively transferred on the acceptor and donor networks). However, other aggregates do not appear as "more negative patches" in the SPV image, and the SPV contrast displays fluctuations over all domains. This observation is also in line with the existence of a

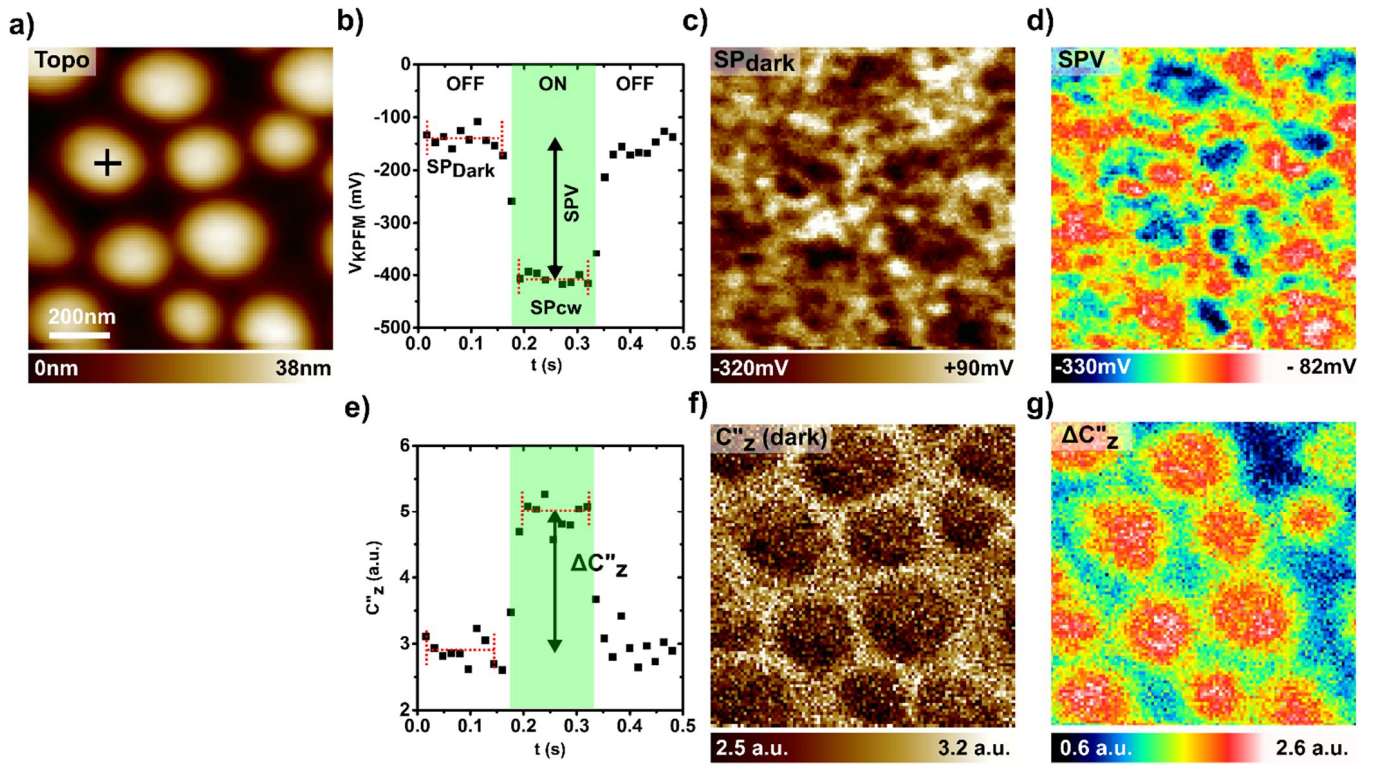

**Figure 6** nc-AFM/KPFM images (1μm×1μm, 100×100pixels) of the PTB7:PC$_{71}$BM sample recorded in data-cube differential SPV imaging mode (λ=515nm, P$_{opt}$=61mW.cm$^{-2}$). **(a)** Topographic image. **(b)** Spectroscopic curve of the KPFM potential as a function of time (30 pixels) recorded during an illumination sequence at the location indicated by a cross in the topographic image. The in dark surface potential (SP$_{dark}$) and surface potential under cw illumination (SP$_{cw}$) are calculated by averaging multiple data points from the curve. **(c,d)** Calculated images of the in dark potential and of the surface photovoltage (SPV=SP$_{cw}$-SP$_{dark}$). **(e,f,g)** Data recorded from the second harmonic channel. **(e)** Capacitance second derivative as a function of time. **(f,g)** Calculated images of C″$_z$ in the dark state, and of C″$_z$ shift induced by illumination.

polymer-enriched skin layer that partially covers the aggregates. A question that might be asked concerns the thickness of the capping layer. Although it is not possible to draw a definite conclusion, complementary data obtained by second harmonic imaging (**Figure 6e-g**) show that it must be thin. In FM-KPFM, it is indeed possible to measure the frequency shift modulation at the double of the electrostatic excitation frequency (along with the first harmonic used as an input for the potential compensation loop), which yields a measurement of the tip-surface capacitance second z-derivative C″z.

Our measurements show that both the "in-dark" and the differential (*i.e.* data under illumination minus data under dark) capacitance signal images display an almost perfect one-to-one correlation with the topographic image. That remarkable behaviour is not so surprising, since the capacitance signal is known to be more sensitive to recessed interfaces; second harmonic KPFM has been

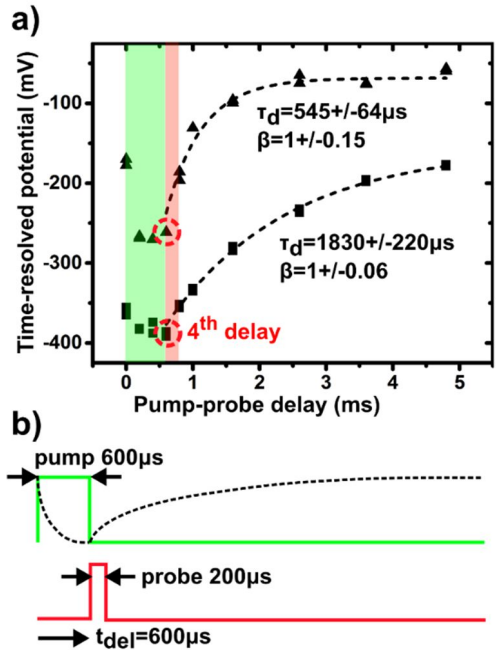

**Figure 7 (a)** Curves of the time-resolved potential probed by pp-KPFM (sum of the dual loops compensation potentials, see the Methods) as a function of the pump-probe delay. For this measurement the pump and probe duration have been set to 600µs and 200µs, respectively (λ=515nm, peak optical power 61mW.cm$^{-2}$). The measurement has been repeated 4 times for each pump-probe delay, only the last two acquisitions have been kept. The time intervals corresponding to the pump and the probe signals (shown only for one given delay) are highlighted by half-transparent green and red rectangles, respectively. The curves have been adjusted in the SPV decay-regime by using Equ. (1) (see text). **(b)** Representation of the pump and probe time-windows specifically implemented for subsurface imaging.[49] Despite this, the absence of intra-domains contrasts in the capacitance signals favours the hypothesis of a very thin skin layer. Otherwise, one shall observe some contrasts over the PC$_{71}$BM aggregates, since the capacitance signal depends on the depth[49] at which they are recessed below the surface.

We now focus the discussion on the blend photo-response probed by pump-probe KPFM. **Figure 7** presents two pp-KPFM spectra acquired at different locations of the sample during a data-cube acquisition. For now, we will not show dynamical images issued from that measurement (the reason for this will become clear in a moment). The pump and probe width were set to 600µs and 200µs over a 5ms repetition period, respectively. Note also that in contrast to the former measurement on the c-Si sample, we made here the choice to use an uneven distribution of pump-probe delays.[23] This allows reducing the data acquisition time, which becomes necessary when performing a data-cube acquisition. At first glance, the delays distribution appears sufficient to analyse the SPV decay after the light pulse extinction. On the other hand, we note that the probe width is too large to track properly the SPV build up dynamics under the optical pump duration.

At first sight, these spectra seem fully consistent with the existence of the negative surface photovoltage, which has previously been observed by "conventional" SPV imaging. By contrast with the complex behaviour observed in the case of the c-Si sample, the time-resolved SP seem to exhibit only a monotonic dependence both under the pump time-window and in between the light pulses. In a first approach, it is tempting to analyse the SPV decay dynamics by adjusting the data with the "single SPV component" formula (Equ.3). This adjustment suggests that SPV decay dynamics display significant variations over the sample's surface, with time constant values ranging from a few hundreds of µs to a few ms.

Actually, mapping a 2D image (shown in the supplementary materials, **Figure S3**) of the decay time constant (calculated by batch processing from the 2D matrix of spectroscopic curves) would reveal that the longer time decays are found over the $PC_{71}BM$ aggregates. This trend is consistent with the existence of electron trapping mechanisms (followed by slow trap-release processes) in the acceptor clusters.

However, one should not too hastily conclude that this scenario encompasses the full physics of the photo-carrier dynamics. A closer examination of the data reveals indeed that the fit displays some deviations from the experimental data for the fourth pump-probe delay of the sequence ($t_{del} = \Delta t = t_0 = 600 \mu s$, **Figure 7b**). This suggests that just after the light pulse extinction, dynamical processes may occur at a faster time-scale than the one defined by the probe time-window.

To verify this, we repeated the measurement with a faster repetition period and a shorter probe time window (note that this scan was performed on a different sample location). In addition, the parameters were also fixed to track more accurately the photocharging dynamics under optical pumping. For that reason, the pump duration was not reduced, which allows recording more data points thanks to the smaller probe time-window.

The results of that second data-cube acquisition are presented in **Figure 8**, which displays two spectroscopic curves recorded on distinct locations of the sample, along with 2D differential maps of the time-resolved potential (*i.e.* images of the differences between the values taken by the time-resolved potential for two different pump-probe delays). Some of the spectroscopic curves look similar to the ones recorded during the previous experiment; they simply reveal the development of a negative photo-potential under illumination. For these curves (referred to as type-I curves), like before, fitting the SPV decay with Equ.3 yields time-constants of a few hundreds of microseconds to a few ms.

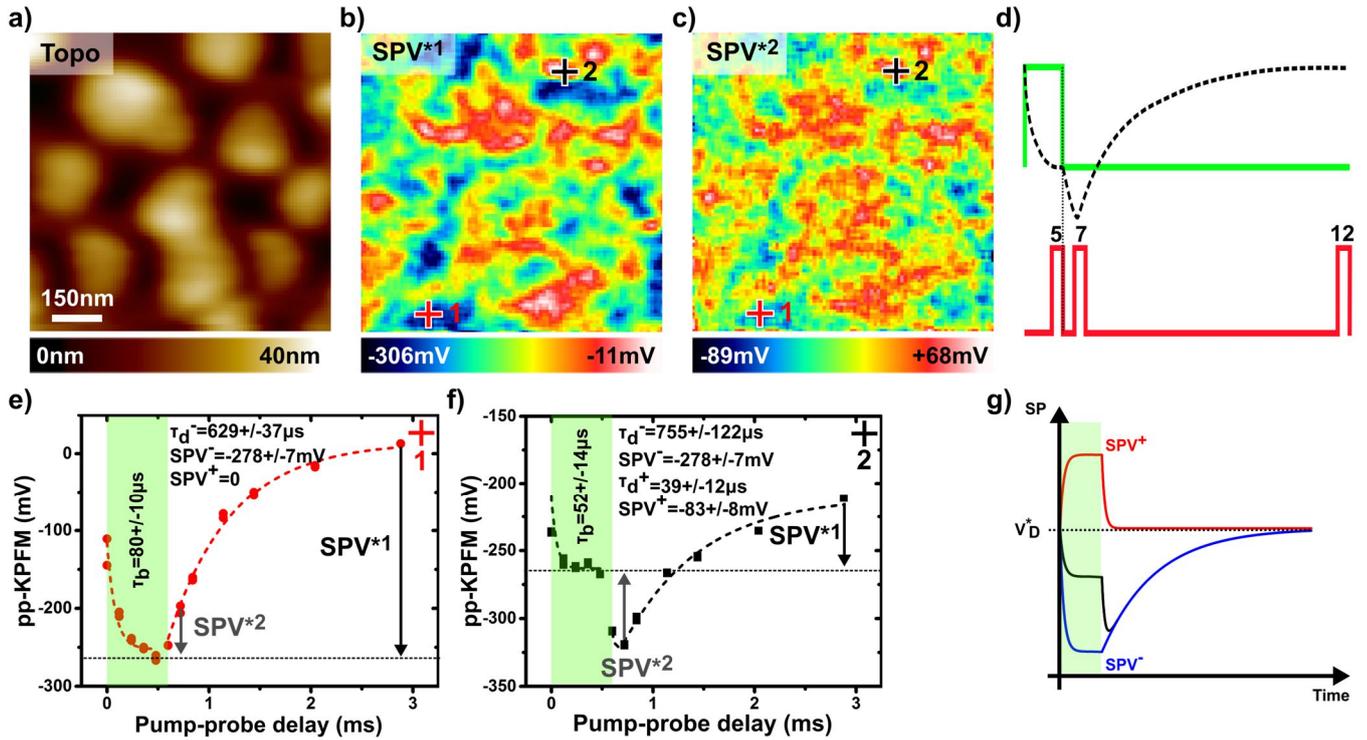

**Figure 8** nc-AFM/KPFM images (900nm×900nm, 90×90pixels) of the PTB7:PC$_{71}$BM sample recorded in data-cube pump-probe KPFM ($\lambda$=515nm, P$_{opt}$=61mW.cm$^{-2}$). Pump duration 600µs. Probe time-window 120µs. Repetition period 3ms. The measurement has been repeated 4 times for each pump-probe delay, only the last two acquisitions have been kept **(a)** Topographic image **(b,c)** "Pseudo SPV" images, obtained by mapping the difference between the signals measured for the 5$^{th}$ and 12$^{th}$ delays (b), and the 7$^{th}$ and 12$^{th}$ delays (c). **(d)** Representation of the pump and probe time-windows as a function of time for the 5$^{th}$, 7$^{th}$ and 12$^{th}$ delays of the sequence. The surface potential is symbolized by a dotted line. **(e,f)** Spectroscopic curves of the time-resolved pp-KPFM potential (sum of the dual loops compensation potentials) as a function of the pump-probe delay. The curves have been recorded on two different location of the surface, highlighted by crosses labeled 1 and 2 in the pseudo SPV images. **(e)** The SPV decay regime can be adjusted by a single-component model (Type I curve, Equ.1). **(f)** A dual component model (Type II curve, Equ. 2) is needed to adjust the SPV decay regime. **(g)** The global SPV can be accounted by taking into account the cumulative contributions of a positive SPV and negative SPV characterized by different decay dynamics.

However, other curves (type II) display a very remarkable feature: just after the light pulse extinction, the time-resolved surface potential drops abruptly to a more negative level than the one reached at the end of the pump light pulse. This drop is followed by a slower return towards the in-dark state.

These anomalous curves are not randomly distributed in the 2D matrix of data; the differential images prove unquestionably that type-II curves are only recorded in specific sample's areas. A "pseudo-SPV" image (**Figure 8b**) has been obtained by mapping the difference between the time-resolved potentials recorded at the end of the light pulse ($\Delta t=480\mu s$, VI) and at the end of the sequence ($\Delta t=2880\mu s$, $V^*_D$). We remind that this difference is referred to as a "pseudo SPV" since the potential cannot fully return to its dark-state level within the sequence period ($V^*_D \leq V_D$). Eventually, mapping the pseudo SPV reveals that the spectroscopic curves display a type-II behaviour in the areas where the photovoltage is less negative. This correlation appears also clearly by mapping the difference between the time-resolved potential values recorded for $\Delta t=480\mu s$ (end of the light pulse) and $\Delta t=720\mu s$ (second delay after the light pulse extinction): above-mentioned areas appear now as "positive patches" in the differential image (**Figure 8c**).

All in all, like in the case of $Al_2O_3$ passivated c-Si samples, a "dual component" photovoltage model is needed to account for the SPV decay dynamics of the bulk heterojunction (**Figure 8g**). Again, both positive and negative components contribute to the global SPV, but the positive SPV magnitude varies depending on the sample location. Where it exists, the positive SPV decays faster after light extinction than its negative counterpart. However, the ratio of the SPV components amplitudes is inverted with respect to the Si-c case: the negative amplitude is larger than the positive one.

Under continuous wave illumination, the SP shift is therefore always negative. An indirect consequence of this change in the SPV components relative weights (with respect to the Si-c case), is that it becomes quite difficult (not to say impossible) to perform a reliable adjustment of the photocharging regime data with the two time-constants model (Equ.7). Specifically, the adjustment cannot be properly performed if all dynamical parameters are kept as free variables ($\tau^+_r$, $SPV^+$, $\tau^-_r$, $SPV^-$). Simply put, the shape of the curves can be roughly accounted either by varying the negative rise time constant, or by increasing (decreasing) the relative weight of the positive SPV amplitude. Obviously, this issue only affects the treatment of the data if there is a non-zero positive SPV amplitude. Type-I curves can be adjusted by using the single-component equation derived to model the SPV dynamics in the photocharging regime, yielding rise time constants of a few tens of μs.

Fortunately, with regards to the decay dynamics, it is possible to adjust all the data (*i.e.* type I or type II curves). The question is how best to assess if a given curve shall be adjusted by using the single or the dual component model. In other words, an objective criterion shall be defined before performing an automated adjustment of the full set of spectroscopic data. Indeed, our goal goes beyond the adjustment of a few selected curves, the data-cube acquisition is first and foremost aimed at mapping 2D dynamical images. To circumvent this problem, a double-pass adjustment procedure has finally been implemented, which made it possible to map the SPV decay dynamics.

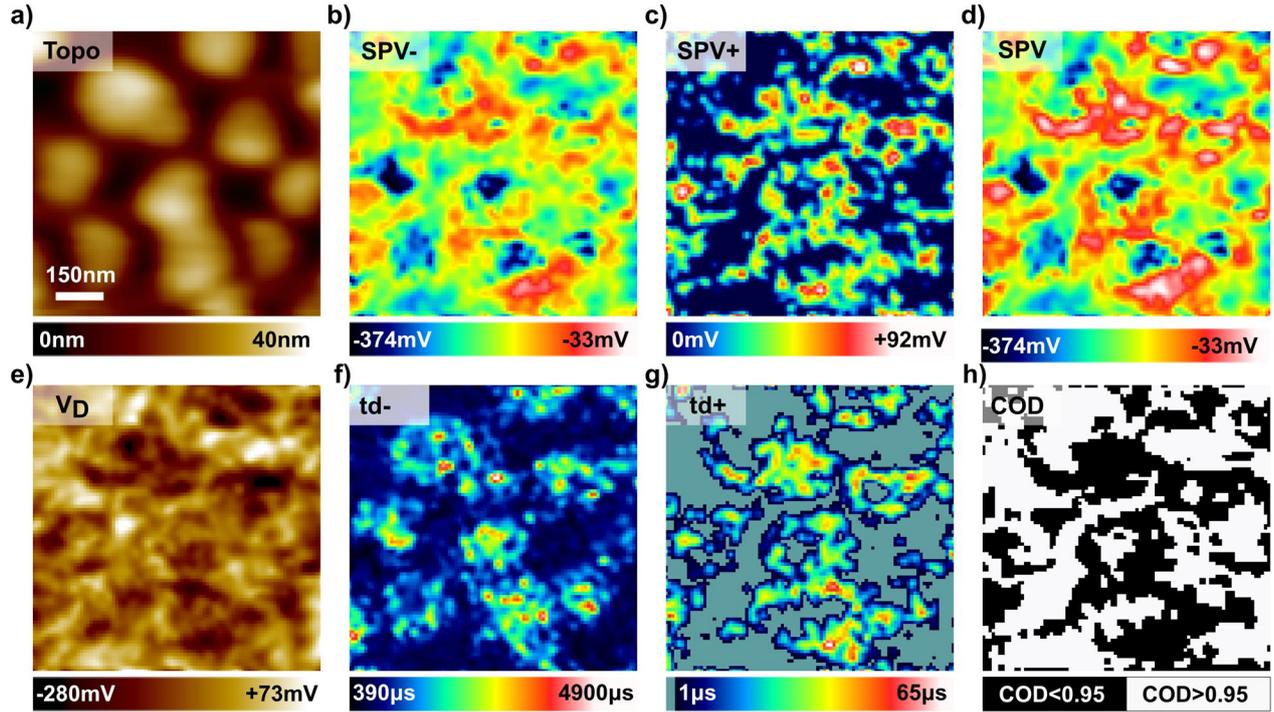

**Figure 9** Dynamical images calculation by batch processing (automated fit) of the 2D matrix of pump-probe curves. **(a)** Topographic image (same data as in Figure 4). **(b,c)** Amplitudes of the negative (SPV-) and positive (SPV+) surface photovoltage components. **(d)** Total surface photovoltage (SPV=SPV$^+$+SPV$^-$). **(e)** In dark surface potential V$_D$. **(f,g)** decay time constants τd$^-$ and τd$^+$ of the negative (f) and positive (g) SPV components. In (g), the uniform gray-green color code (at the left of the color scale) correspond to the areas where SPV$^+$=0 (no τd$^+$ adjustment). **(h)** 2D map of the coefficient of determination calculated when performing an adjustment based on a single component model (Equ. 3, no positive SPV component). The images have been calculated by merging the outputs of two successive adjustments with the single component and dual component models (see the supplementary information).

First, a Gaussian smooth was applied to the raw spectroscopic data matrix (3 adjacent pixels) to increase the signal to noise ratio (dynamical images obtained from the raw data display similar features, but with a higher signal to noise ratio). The data were subsequently adjusted a first time by using the "single component" decay equation (Equ 3) with a negative SPV amplitude (Equ 3).

This procedure allows identifying areas which can be properly characterized by a unique negative SPV component (i.e. curves displaying a type I behaviour). To do so, the following criterion is applied. The parameters extracted from that first adjustment will be stored if the coefficient of determination (COD) is higher or equal than 0.95, in that case a zero magnitude is a posteriori ascribed to the positive SPV component (*i.e.* SPV$^+$=0, no τ$_d^+$ ). For the sake of completeness, it should be noted that this process has been repeated a few times with different COD thresholds. We found empirically that fixing the threshold to 0.95 allows a proper identification of "type-I areas". This appears clearly when comparing a 2D map of the COD and the pseudo SPV images (this empirical part of the process could however probably be further optimized in future works).

A second fit is subsequently performed, during which the data are now adjusted by using the double-component decay model (Equ. 6). The outputs of the first and second fits are combined to map a single set of 2D dynamical images ($\tau^+_d$, $SPV^+$, $\tau^-_d$, $SPV^-$). The merging is simply achieved by ascribing to each image pixel the output value from the first or second adjustment (the attribution being done as a function of the COD value calculated during the first fit). The flow chart of the double-pass adjustment is detailed in the supplementary information (**Figure S4**).

The complete set of 2D dynamical images is presented in **Figure 9** along with the topographic data, the in dark surface potential ($V_D$) and the 2D map of the COD used for the two-pass adjustment procedure. Note that the $V_D$ image is also an output if the adjustment procedure. One can see that the contrasts in the SPV component amplitudes images are fully consistent with the ones of the "pseudo SPV" images (compare **Figure 9** and **Figure 8**). This confirms that the double-pass adjustment procedure successfully allowed identifying the areas where there is a non-negligible positive SPV component. As expected from our previous analysis on a few selected curves (**Figure 8f**), the positive SPV decays much faster than its negative counterpart, with decay time constants ($\tau^+_d$) on the order a few tens of µs. The negative component decay time constant ($\tau^-_d$) image is remarkably correlated with the topographic image: higher time constants are observed over the mesoscopic $PC_{71}BM$ domains. This last observation confirms our first assumption: the acceptor aggregates act as electron trapping centres; the longer decays originate from slow trap-release processes.

To conclude this part of the work, we need to explain what is the origin of the positive SPV component, and why it decays much faster than the negative one. As explained before, for this kind of PTB7:$PC_{71}BM$ blend (*i.e.* processed without additives), it is now widely admitted that the top skin layer is formed by a donor-acceptor phase which is finely-mixed and polymer-enriched. As a consequence, specific donor-acceptor interfaces are formed when the mesoscopic (and pure) $PC_{71}BM$ aggregates are covered by this PTB7-enriched skin layer. Here, a less negative SPV will be probed by KPFM, since the top part of the film is formed by a phase enriched in the electron donor material (which becomes positively charged under illumination). After the light pulse extinction, the photo-carrier located on both side of these interfaces will experience a fast recombination. At a longer time-scale, electrons that have been trapped in the $PC_{71}BM$ clusters will be released (and will eventually meet a counter charge), accounting for the slow "negative SPV" decay dynamics. Note that in between the mescoscopic aggregates, smaller (*i.e.* nanometer-sized) $PC_{71}BM$ clusters dispersed in the finely mixed D-A phase behave also as electron trapping centres. Therefore, in overall, the negative decay time constant is much higher than the positive one. This last hypothesis is reinforced by the comparison with the data acquired on the reference nano-phase segregated sample processed with DIO additive. In the case of this last sample, indeed, the pp-KPFM curves can be well accounted by using a single-component (negative) SPV, with decay time-constants in average of a few hundreds of µs. Specific contrasts in the dynamical images (see **Figure S5** in the supplementary information) suggest that trapping mechanisms are also at play.

Evidently, the above model is only a first sketch on the way of a better understanding of the photo-carrier dynamics in this kind of bulk heterojunction thin film. We do not claim that all the dynamical processes have been accounted for, and some uncertainties remain with regards to the true nature of the sample morphology and phase composition. In particular, the composition of the skin layer remains an open question. For instance, it is unclear whether the skin layer can in places be formed by pure polymer material or not. Further insights on these questions may be gained by investigating planar interfaces[50] realized by evaporating acceptor molecules (such as C60 or C70) on pure PTB7 films.

## Conclusion and remarks

To summarize, performing pump-probe KPFM measurements on $Al_2O_3$-passivated Si-c and PTB7:$PC_{71}$BM blends allowed us to reveal the existence of light-induced charge distributions that would have been hidden to conventional SPV imaging. There is a strong likelihood that similar situations occur in other inorganic, organic and hybrid systems. This urge us to apply more systematically time-resolved KPFM modes to local investigations of photovoltaic samples. In hindsight, one can know realize how great the risk is to misinterpret data acquired by conventional SPV imaging on photoactive samples with an heterogeneous composition or morphology. For instance, by assuming that the SPV yields a local measurement of the open circuit voltage ($V_{OC}$), the smaller magnitude of the signal observed over some areas of the PTB7:PC71BM blend could have been ascribed to a higher recombination rate.[51] Such interpretation follows indeed naturally the well-known dependency of the quasi Fermi levels splitting with respect to the photo-carrier recombination rate. While this picture of a local $V_{OC}$ measurement may still be relevant in some cases, our work show that it shall not lightly be applied without considering the real nature of the internal sample charge distributions within the sample.

Our results highlight also the usefulness of implementing techniques that can provide a quantitative measurement of the time-resolved potential. We note that it may be more difficult to come to the same conclusions (about the nature of the sample photo-induced charge distributions) when using time-resolved techniques based on the analysis of photocapacitive signals. In that sense, G-mode KPFM[13] appears as an interesting alternative to pp-KPFM.

It is also important to keep in mind that some SPV components may remain hidden to pp-KPFM investigations, for instance because some of the underlying dynamical process occur at much shorter time-scales that the probe time-window. In principle, the resolution of pump-probe KPFM can be pushed down the sub-10ns time-scale, which holds out the promise to investigate fast recombination processes. However, these fast processes will be revealed only if the contributions of charge populations which slowly evolve in time remain negligible. For that reason, it may be in general highly relevant to compare the outputs of pp-KPFM measurements with the ones of other techniques, such as time-resolved photoluminescence.

Despite these remaining difficulties, we must not lose sight of the progress that have been made possible these past years thanks to the advent of time-resolved KPFM. Our work demonstrated that in addition to proving an access to the SPV dynamics, pp-KPFM can be used to perform a differentiated analysis of the photo-induced charge distributions within a sample. We anticipate that similar effects than the ones reported herein will be observed in other kinds of opto-electronic and photovoltaic systems, such as type-II van der Waals heterostructures processed on dielectric substrates.

## Authors contributions

B.G. implemented the pp-KPFM setup, carried out the experiments on the organic blends, and analysed the data obtained on these samples. O.B. processed the PTB7:PC$_{71}$BM blends under the supervision of R.D. K.C. provided the Al$_2$O$_3$-passivated c-Si sample. V. A. performed the experiments on c-Si and analysed the data under the supervision of Ł.B. B.G. wrote the manuscript, with inputs from V.A. for the c-Si part.

## Acknowledgments

Part of this work, carried out on the Platform for Nanocharacterisation (PFNC), was supported by the "Recherches Technologiques de Base" program and Carnot funding of the French National Research Agency (ANR).

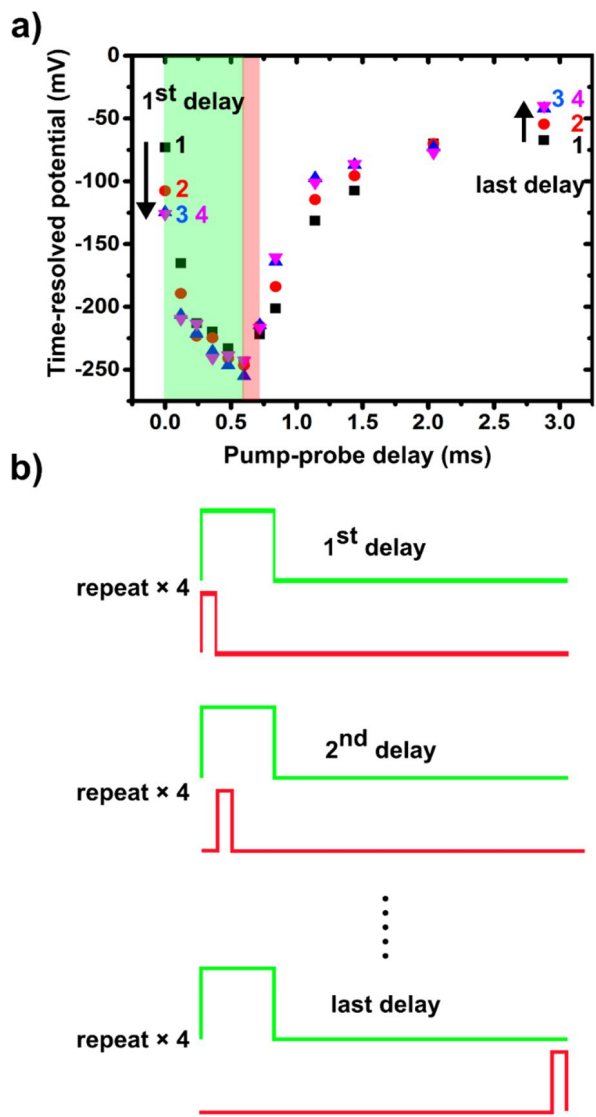

**Figure S1** Curve of the time-resolved KPFM potential (sum of the dual compensation potentials, see the methods in the main text) as a function of the pump-probe delay. Data acquired on the $PTB_7:PC_{71}BM$ sample with a pump duration of 600μs and a probe time-window of 120μs (sequence period : 3ms). λ=515nm, $P_{opt}$=61mW.cm$^{-2}$. Four data pixels are successively acquired for each pump-probe delay. Only the two last delays are kept for the post-acquisition data analysis. This procedure ensures that enough time is given to the pp-KPFM feedback loop to track the changes that occur after shifting the pump-probe delay to a new value.

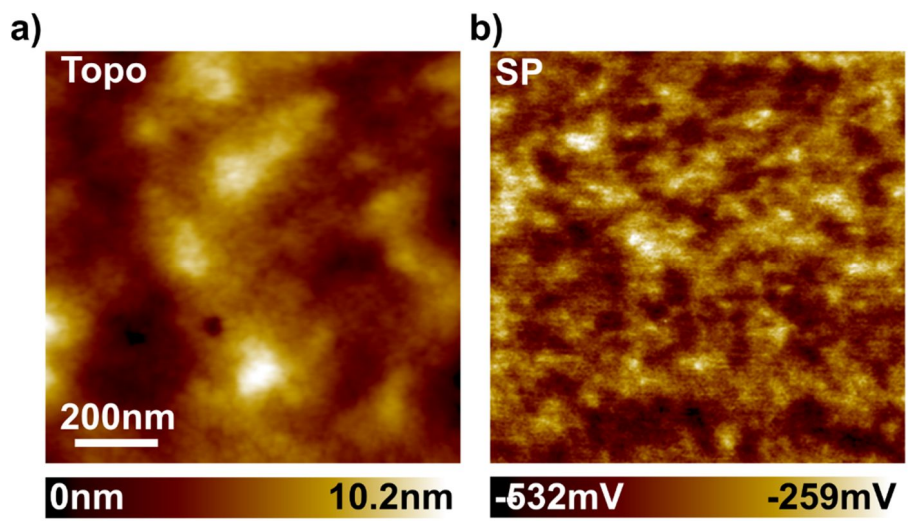

**Figure S2** nc-AFM/KPFM images (1μm×1μm, 300×300 pixels) of the reference PTB7:PC$_{71}$BM blend processed from a CB solution with a DIO additive. (a) Topography. (b) Surface potential (in dark conditions).

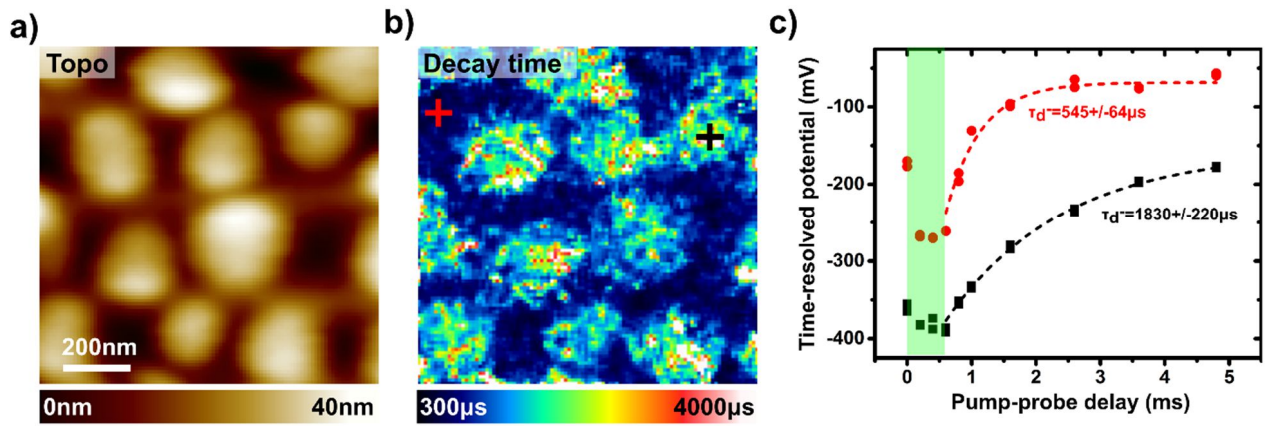

**Figure S3** pp-KPFM spectroscopic imaging of the PTB7:PC$_{71}$BM sample processed without DIO with the first pump-probe delay sequence (see main text). Pump duration : 600μs. Probe time-window : 200μs. λ=515nm, P$_{opt}$=61mW.cm$^{-2}$ **(a)** Topographic image **(b)** Decay time-constant image calculated by batch processing with Equ. 1. (the stretched exponent has been fixed to 1) **(c)** pp-KPFM curves recorded at two distinct locations, indicated by crosses in b). Longer decays are observed over the PC$_{71}$BM aggregates.

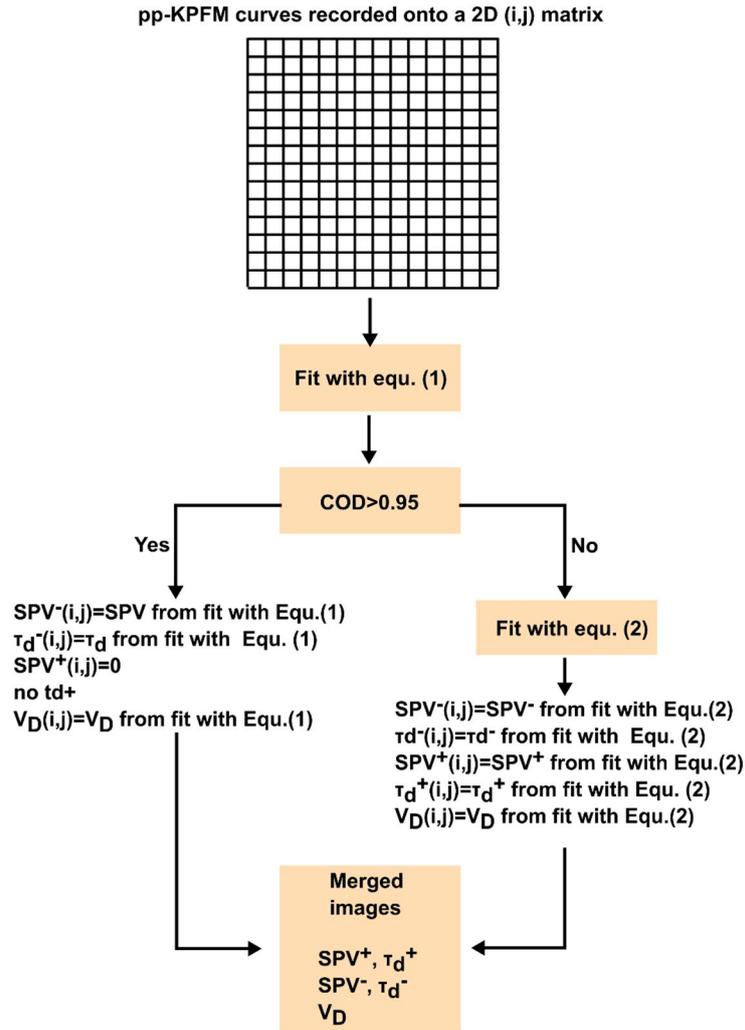

**Figure S4** Dynamical images calculation flowchart. Each curve (i,j) from the 2D data-cube matrix is adjusted with Equ.1. If the coefficient of determination (COD) is equal or above 0.95, the outputs of the fit are kept. In that case the SPV is modelled by a single negative component, a zero magnitude is ascribed to the positive SPV magnitude and there is no related time-decay (i.e. SPV$^+$=0, no $\tau_d^+$). If the COD is lower than 0.95, the (i,j) curve is adjusted by using Equ.2 with a double set of dynamical parameters (SPV$^-$, $\tau_d^-$ and SPV$^+$, $\tau_d^+$). 2D images of the static ($V_D$) and dynamic (SPV$^-$, $\tau_d^-$, SPV$^+$, $\tau_d^+$) parameters are eventually calculated by merging the outputs of both fits.

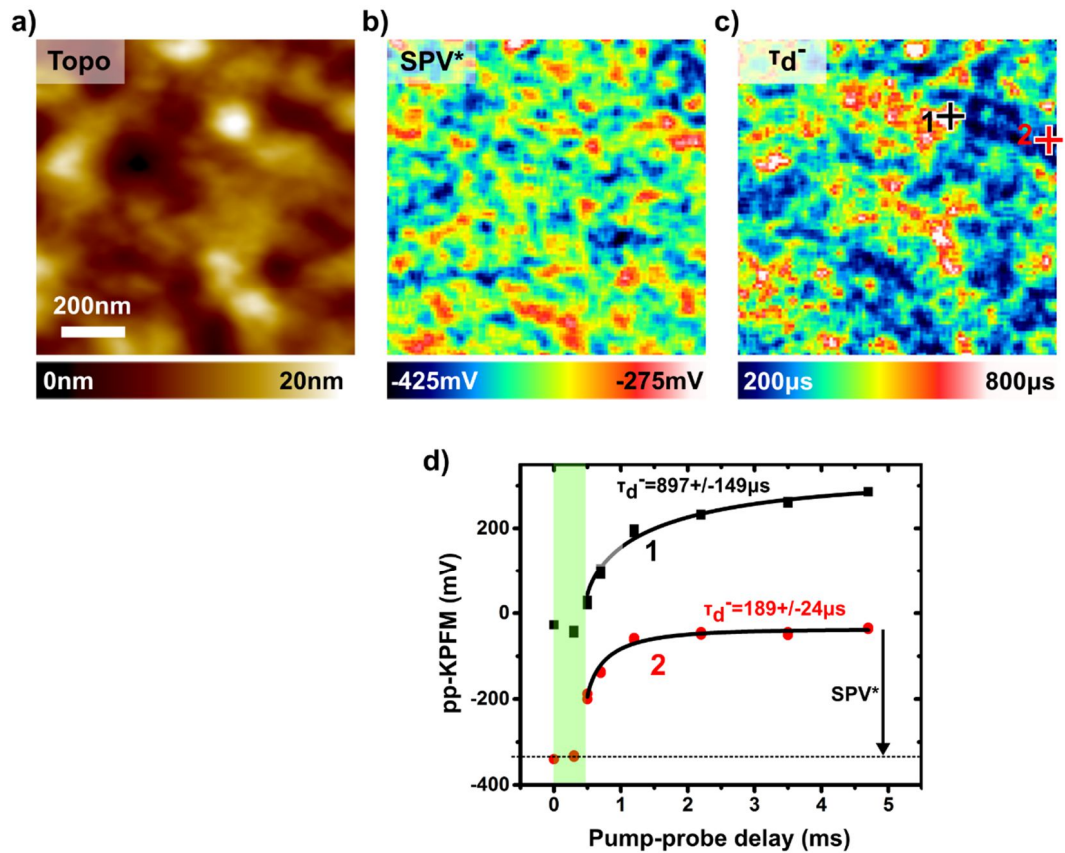

**Figure S5** pp-KPFM spectroscopic imaging (1000nm×1000nm, 100×100pixels) of the PTB7:PC$_{71}$BM reference sample processed with DIO. Pump duration : 500μs. Probe time-window : 200μs. λ=515nm, P$_{opt}$=77mW.cm$^{-2}$ **(a)** Topographic image **(b)** Pseudo SPV image obtained by mapping the difference between the signals measured for the 2$^{nd}$ and 8$^{th}$ delays. **(c)** Decay time-constant image calculated by batch processing with Equ. 3. (see main text) **(d)** pp-KPFM curves recorded at two distinct locations, indicated by crosses in c). Longer decays observed over some areas reveal the existence of electron trapping centres.